*DK-PRACTICE: An Intelligent Educational Platform for Personalized Learning Content Recommendations Based on Students Knowledge State*

Marina Delianidi, International Hellenic University, Greece
Konstantinos Diamantaras, International Hellenic University, Greece
Ioannis Moras International Hellenic University, Greece
Antonis Sidiropoulos, International Hellenic University, Greece

**Abstract**
This study introduces DK-PRACTICE (Dynamic Knowledge Prediction and Educational Content Recommendation System), an intelligent online platform that leverages machine learning to provide personalized learning recommendations based on student knowledge state. Students participate in a short, adaptive assessment using the question-and-answer method regarding key concepts in a specific knowledge domain. The system dynamically selects the next question for each student based on the correctness and accuracy of their previous answers. After the test is completed, DK-PRACTICE analyzes students' interaction history to recommend learning materials to empower the student's knowledge state in identified knowledge gaps. Both question selection and learning material recommendations are based on machine learning models trained using anonymized data from a real learning environment. To provide self-assessment and monitor learning progress, DK-PRACTICE allows students to take two tests: one pre-teaching and one post-teaching. After each test, a report is generated with detailed results. In addition, the platform offers functions to visualize learning progress based on recorded test statistics. DK-PRACTICE promotes adaptive and personalized learning by empowering students with self-assessment capabilities and providing instructors with valuable information about students' knowledge levels. DK-PRACTICE can be extended to various educational environments and knowledge domains, provided the necessary data is available according to the educational topics. A subsequent paper will present the methodology for the experimental application and evaluation of the platform.

# Introduction

Learning is a fundamental aspect of human activity from birth onward, serving as the foundation for cultivating and expanding knowledge. In the educational process, knowledge is developed by engaging with various knowledge elements through appropriate educational materials. Educational technology enhances this process by employing strategies and tools to optimize learning and performance. These tools are applied in various educational settings, including formal, informal, or lifelong learning environments (Huang et al., 2019). Knowledge evolves over time, and this dynamic nature has been a key focus of research in the educational data mining field in recent years. Specifically, a student's knowledge improves when they acquire a new learning object such as a concept or skill. Maintaining, over time, the probability that the student has mastered various learning objects is called Knowledge Tracing (KT) (Corbett and Anderson, 1994). KT research first emerged in the 1990s, utilizing Bayesian models (Corbett & Anderson, 1994). This line of research has evolved, aligning with modern technologies and incorporating Machine Learning (Cen et al., 2006; Pavlik Jr et al.,2009) as well as Deep Neural Network techniques such as Recurrent Neural Networks (Piech et al., 2015; Zhang et al., 2017; Yeung, 2019; Delianidi et al., 2021), Convolutional Neural Networks (Yang S. et al., 2020; Wang et al., 2020; Shen et al., 2020; Ma et al., 2021), or Graph Neural Networks (Yang Y.et al., 2020; He et al., 2021).

Estimating students' performance enables an evaluation of their knowledge state and supports the enhancement of their future learning outcomes by tracing their current knowledge. In the context of learning with an Intelligent Tutoring System (ITS), students' interactions are automatically logged as they interact with the system. These log files can be analyzed during the KT process to monitor how the student's knowledge state evolves over time. So, e-learning constitutes a vital component of educational technology. Especially during the COVID-19, e-learning has become the primary global educational tool, creating unprecedented opportunities for learning beyond the traditional confines of time and space (Alqahtani & Rajkhan, 2020). Through e-learning, learners can access digital materials such as presentations, texts, videos, audio, and images from various sources, including structured courses on learning management systems, teacher websites, educational social networks, e-books, and online repositories. To further enhance this learning experience, there is a growing need for smart applications that cater to individual student needs by delivering personalized educational content within online learning environments.

One example of such intelligent applications is Educational Recommendation Systems. These systems offer personalized suggestions to tailor learning experiences to individual students' needs (Hukkeri & Goudar, 2022). In today's era of digital learning and information overload, personalization plays a critical role in improving students' performance and knowledge evolution. Advances in machine learning and recommendation technologies have opened up promising opportunities for education. Their use enhances learning for students, instruction for teachers, and resource management for institutions (Zou, 2011). Beyond traditional education, these systems also support lifelong learning, addressing the increasing demand for ongoing skill development in modern society. Educational recommendation systems are reforming traditional learning by fostering innovation in teaching and educational practices (Chen & Yang, 2023). The integration of recommendation systems into education aims to assist students and teachers in multiple ways. For instance, by providing users with relevant resources - such as articles, books, videos, and courses - these systems help save time and effort in searching for information while enabling the discovery of new resources that might otherwise be overlooked (Belarbi et al., 2019; Kim et al., 2019; Gulzar et al., 2018). Additionally, recommendations based on users' past behaviors, such as courses they've taken (Shin & Bulut, 2022), or preferences, like learning styles or interests (Giang et al., 2023), not only enhance accuracy but also make it easier to locate necessary information. This is particularly valuable in online learning, where vast amounts of information are dispersed across the internet.

Drawing inspiration from research in knowledge tracing and educational recommendations, in this paper we introduce the DK-PRACTICE, an intelligent platform designed to support adaptive learning and deliver tailored educational recommendations to learners, utilizing machine learning and deep learning methods. Based on Machine Learning and Deep Learning methods, the platform combines the knowledge tracing task and targeted educational content recommendation after first discovering the learner's knowledge gaps through their participation in a short test.

**Main Purpose**

The DK-PRACTICE platform is an intelligent system designed to deploy personalized educational recommendations, utilizing machine and deep learning to create an adaptive learning environment tailored to each student's needs. Starting with an initial assessment to identify knowledge gaps, it recommends targeted learning materials. This process includes a brief interactive placement test and the analysis of students' past question-and-answer interactions. DK-PRACTICE aims to strengthen students' foundational knowledge, offering self-assessment opportunities and supporting exam preparation. Tutors can also quickly gauge students' knowledge levels, enhancing teaching effectiveness. The determination of students' personalized educational requisites is initiated through a succinct personalized interactive knowledge placement test, supplemented by analysis of students' historical interactions. Leveraging this information, the DK-PRACTICE platform generates personalized recommendations aimed at enriching the student's knowledge and academic achievements. Beyond the platform's development, our principal aim is to fortify students' educational fundamentals through an interactional process with the platform. By detecting knowledge gaps, the DK-PRACTICE platform facilitates the provision of suitable educational content for the students' knowledge remediation. The creation of this platform is motivated by several factors, notably affording students the opportunity to assess their knowledge, particularly in anticipation of examinations. Moreover, DK-PRACTICE aims to equip tutors with an expeditious means to gauge the cognitive proficiency levels of their students, thereby enhancing pedagogical efficacy. In summary, the main objectives of the DK-PRACTICE platform are:
- to improve the performance of students based on their past interactions (i.e. responses to test questions),
- to dynamically recommend appropriate, personalized educational content in a specific subject of study,
- to explore the student's knowledge before a test or as an auxiliary tool for the tutors to know the knowledge level of their students.

**DK-PRACTICE Architecture**

The architecture of the DK-PRACTICE platform includes two main components (Fig.1):
- The first component combines a recurrent Deep Learning (DL) model with a classification sub-network to estimate the student's knowledge state (Fig.2).
- The second component applies Machine Learning (ML) algorithms in order to produce educational content recommendations (Fig.3).

Both, the Knowledge Tracing task and the Knowledge State estimation are accomplished by utilizing the student's interaction history.

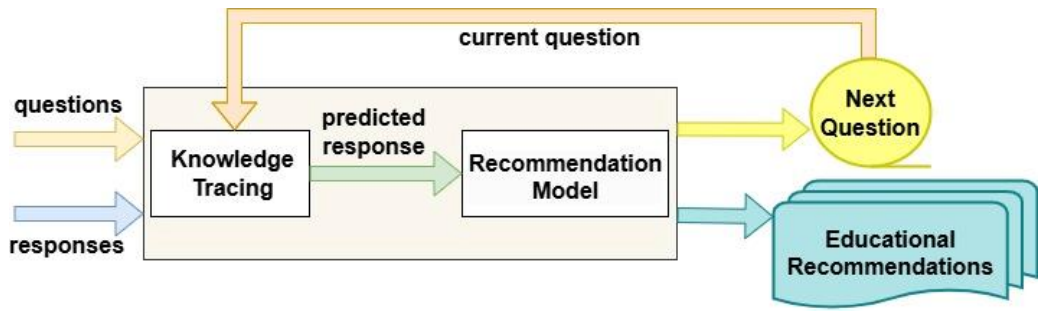

Figure 1: The DK-PRACTICE architecture.

**Knowledge Tracing Model**

The knowledge tracing model is based on previously published work (Delianidi & Diamantaras, 2023). As depicted in Figure 2, the interaction involves administering a question-and-answer test, with initially encoded questions and responses. The historical record of question-answer pairs serves as input to the DL model to estimate the student's knowledge state. This estimated knowledge representation, alongside a new, previously unseen question, is fed into a classification network. The network then calculates the probability of the student's correct response, referred to as the "predicted response" $r'$. The "predicted response" is the element that affects the choice of the next question $q$, which is different for each student during the test. It contributes to the list of content recommendations after completing the test.

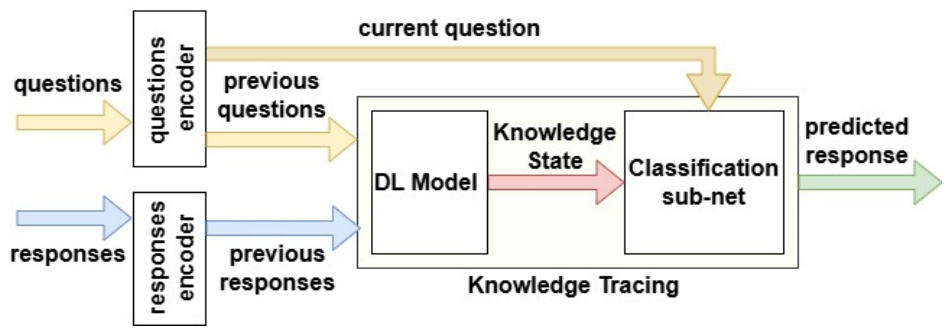

Figure 2: The Knowledge Tracing model.

**Recommendation Approach**

The Recommendation Engine (Fig. 3) produces two types of recommendations:
   a) the *"Next question"* and
   b) the *"Educational Recommendations"*.

There is a question bank where each question concerns a knowledge component and arises from a specific concept. Initially, the system selects a random question. Depending on the student's response, it estimates the probability of answering correctly all the questions in the question bank. Drawing inspiration from the NeurIPS 2020 Education Challenge[1], we define

---
[1] NeurIPS 2020 Education Challenge, https://eedi.com/projects/neurips-education-challenge , last accessed 2024/01/03

an ambiguity threshold $\alpha$, the value of which is compared to the probability of answering correctly to the questions. The higher the probability of the "predicted response" $r'$, which was the output of the knowledge tracing component, the better the student's knowledge of the concept to which the question refers. In the case where the probability of the "predicted response" $r'$ is below a certain threshold $\alpha$, the system believes that the student does not sufficiently understand the concepts referred to in the question. On the other hand, the smaller the difference between $\alpha$ and the estimated probability of correctness $r'$ for each question $q$, the greater the system's doubt about the student's knowledge state per concept.

$$\text{If } r'_q \cong \alpha, \text{ then "}Next\ question\text{"} = q$$

Consequently, guided by the prediction of the DL model, the question $q$ chosen is the one that has the predicted probability $r'$ approaching $\alpha$, where *α=0.5*. The reason for this choice is to prompt the student to answer a question which the model has the most difficulty to classify as likely to be answered correctly or not. In any case, the *"Next question"* recommendation differs for each student and depends on their current performance during the short test interactions.

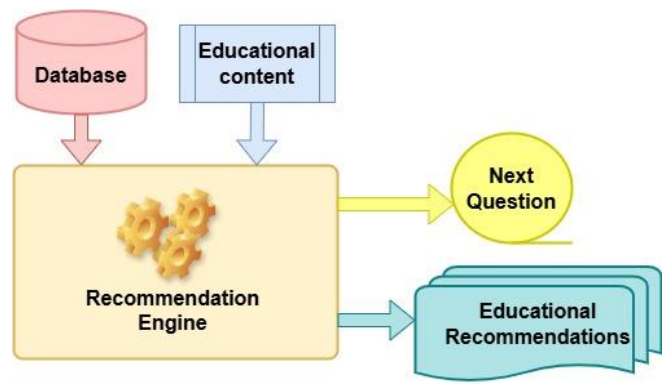

Figure 3: The Recommendation Model

Following a sequence of interactions between the student and the platform, wherein each student responds to personalized questions, the system identifies the student's areas of knowledge deficiency. Subsequently, it proposes educational materials aimed at addressing these knowledge gaps. Within the framework of the DK-PRACTICE platform, we have established the personalized question ***quota*** at ***10***. This numerical threshold constitutes a hyperparameter and is subject to occasional adjustments following preferences, experimental findings, course requirements, and similar factors. Based on the student's responses, the recommendation engine is tasked with recommending pertinent educational content conducive to enhancing the student's knowledge proficiency. All this process is supported by a data model which is organized into a database, a training dataset, and a database of educational content. Finally, the recommended educational content is related to the educational concepts that the student has to learn.

**Data Model**

Data is one of the most fundamental components of the DK-PRACTICE platform. It is essential for the operation of any type of application that uses machine learning. The data model of the platform is depicted in Figure 4. The database contains the entities for the questions, the educational concepts, the information about the educational content as well as the relationships between the entities. The dataset consists of the interaction of questions and answers history and is used to train the machine learning models. The data related to the Exams is used to train the Knowledge Trace Model after anonymization.

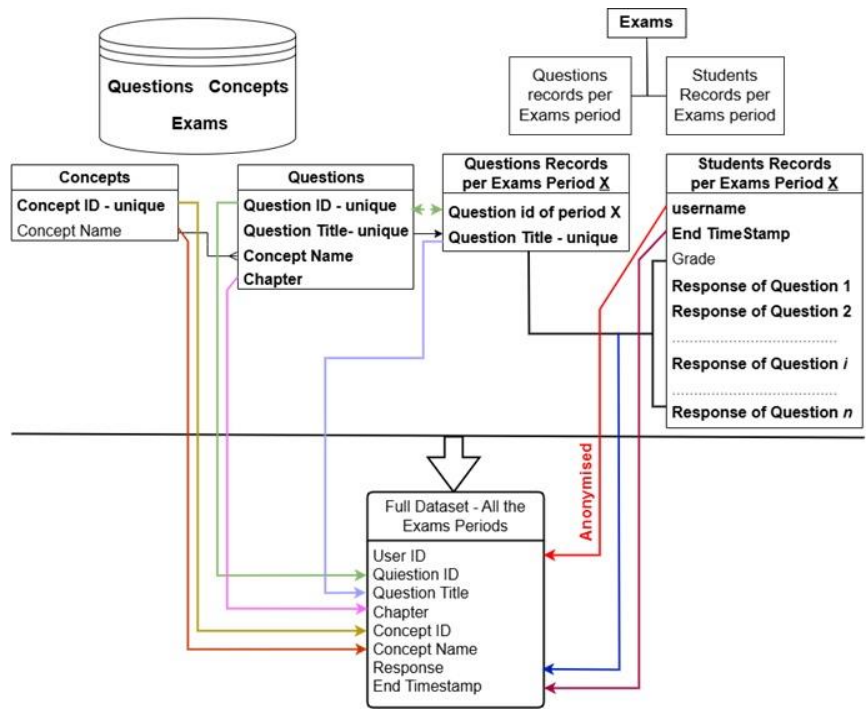

Figure 4: Data Model

**Implementation**

The DK-PRACTICE platform was developed using Laravel[2] framework, one of the most popular PHP frameworks for developing web applications. A major advantage of Laravel is its extensive support for integrating external packages, allowing developers to easily extend the capabilities of their applications. The application connects to an API (Application Programming Interface), implemented in Flask[3], and supports two-way communication with the machine learning models, as the *"Next question"* suggestion and the final list of *"Educational Recommendations",* i.e. educational content recommendations are sent through the API as it enables the Machine Learning model to be leveraged through the developed application.

The DK-PRACTICE platform allows two types of users: the administrator and the student. The administrator has the right to create a test by entering all the relevant information and to set parameters such as the number of test questions and tests a student can take. The administrator can also monitor statistics and the performance of the students. Figures 5-10 present the platform's administrator environment functionalities.

---

[2] Laravel - Wikipedia, En.wikipedia.org, https://en.wikipedia.org/wiki/Laravel , last accessed 2023/12/12
[3] Welcome to Flask - Flask documentation, https://flask.palletsprojects.com/en/3.0.x/ ,last accessed 2024/03/02

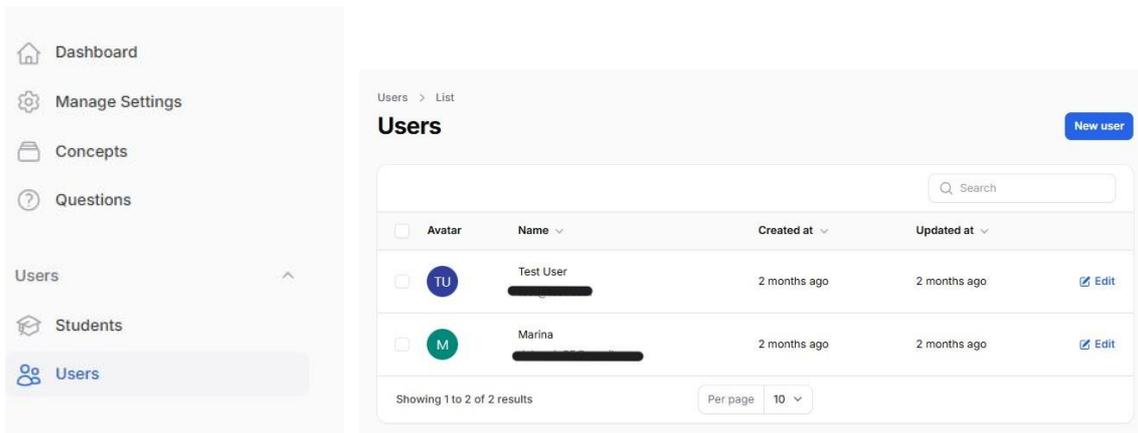

Figure 5: Administrator environment - Users

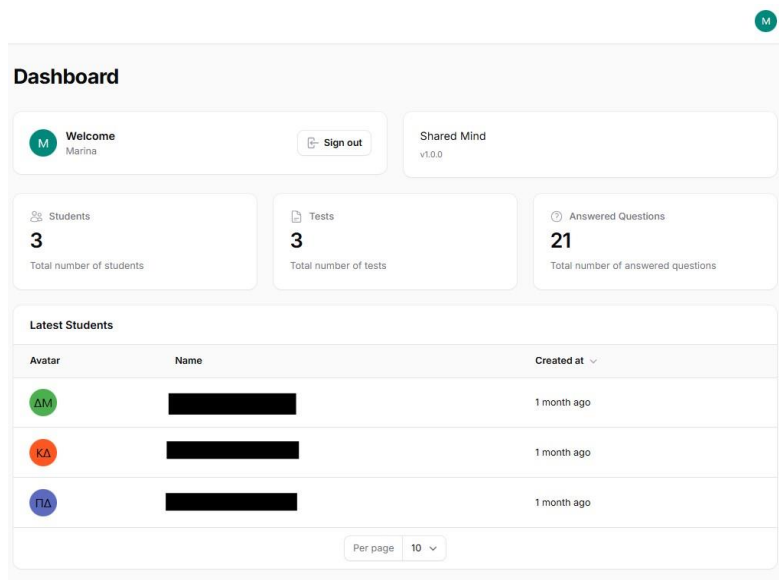

Figure 6: Administrator environment - Students

Figure 7: Administrator environment – Edit student

Figure 8: Administrator environment – The Concepts

Figure 9: Administrator environment – The Questions bank[4]

The student has the right to perform the permitted number of tests, to see the correctness of the answer to the question. By default, the number is set to 2 tests: the first to identify the student's knowledge gaps after completing the test, and the second, after receiving and studying the list of educational content recommended by the platform. Statistics are presented in both administrator and student environments (Figures 10, 11).

Figure 10. Student's tests list

---

[4] The questions are in Greek because the course is taught in Greek.

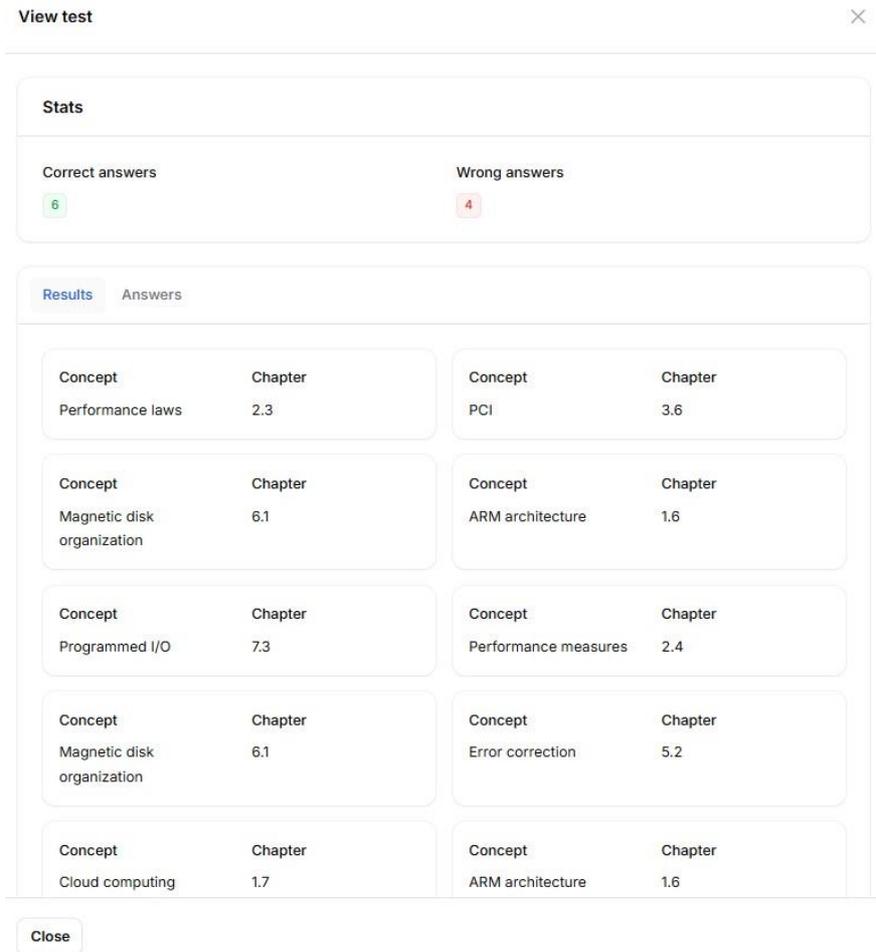

Figure 11. Statistics and Recommendations according to test's results

**Discussion and Future Work**

The DK-PRACTICE platform has undergone preliminary testing within the context of the "Computers Organization and Architecture" course. These initial trials aim to evaluate the platform's technical functionality, identify bugs, and ensure a seamless user experience in preparation for broader deployment. Early feedback indicates that the platform effectively integrates knowledge tracing and personalized recommendations, supporting adaptive learning through data-driven methodologies. The pilot deployment has also provided valuable insights for optimizing the knowledge tracing model, which currently achieves an accuracy of 78.11% and an AUC (Area Under Curve) of 80.49%, using anonymized historical data from previous exam cycles. Moving forward, several key initiatives are planned to improve and expand the platform. A critical next step includes evaluating the educational effectiveness of the platform's recommendations. This will be achieved through a pilot operation by groups of students to prepare for the final exams of the course a few weeks in advance. This will allow for the evaluation of the recommendations of the proposed educational material, the impact and retention of knowledge, and the effectiveness of academic performance. In order to collect comprehensive feedback from users, we plan to distribute a detailed questionnaire. This survey will record students' perceptions of the usability of the platform, the accuracy of knowledge state assessments, and the relevance of content suggestions. According to user feedback, the system can be improved to better meet the needs of students.

Another point is to expand the scope of testing to include diverse courses and larger student groups. Such testing will provide more datasets, enabling further improvement of the machine learning models underlying the knowledge tracing and recommendation components. We also aim to explore incorporating additional educational domains, which would demonstrate the platform's adaptability and scalability. Finally, we plan to extend the platform's functionality beyond academic contexts to professional training environments. By using the relevant datasets and adapting the knowledge tracing and recommendation systems to identify and address skill gaps in employees, the platform could serve as a versatile tool for continuous professional development. With ongoing improvements, DK-PRACTICE has the potential to become a transformative educational technology, supporting personalized and effective learning experiences across a wide range of domains.

**Conclusion**

This paper introduces DK-PRACTICE, an intelligent educational platform designed to facilitate adaptive and personalized learning through educational recommendations based on students' knowledge states. By leveraging machine learning and deep learning techniques, DK-PRACTICE offers an innovative approach to enhancing the learning experience. The platform integrates knowledge tracing to assess individual knowledge gaps and uses these insights to recommend targeted educational content tailored to each student's specific knowledge needs. It also provides instructors with valuable information to understand and address their students' knowledge deficiencies. One of the platform's useful features is its dynamic assessment mechanism, which enables students to undergo short, interactive tests that identify knowledge gaps. The content recommendations, generated by the machine learning components, focus on these gaps, ensuring a targeted and efficient learning process. The inclusion of pre- and post-tests, along with detailed performance reports, allows students to monitor their progress.

The adaptability of DK-PRACTICE is another significant advantage. The platform's architecture supports extension across different educational subjects and levels, making it versatile for various contexts. Additionally, its potential applications extend beyond traditional education settings to professional training environments, where it could play a pivotal role in identifying and addressing skill gaps among employees. By tailoring recommendations to individual learning requirements, DK-PRACTICE can contribute to lifelong learning and workforce development initiatives. However, challenges remain in fully realizing the platform's potential. Technical improvements, such as enhancing the robustness of the recommendation engine and integrating more sophisticated deep learning models, are critical next steps. In conclusion, DK-PRACTICE represents a promising step forward in personalized educational technology. By combining advanced machine learning methodologies with a user-centered design, the platform demonstrates its capacity to transform traditional learning paradigms. Future developments will focus on refining its capabilities and expanding its applicability, aiming to establish DK-PRACTICE as an intelligent platform for adaptive and personalized education for students, tutors, and professionals.

**Contact email:** dmarina@ihu.gr